\documentclass[pre,preprint,superscriptaddress]{revtex4}

\usepackage{graphicx}
\usepackage{color}
\usepackage{amsmath,amssymb}
\usepackage{mathrsfs}

 \newcommand{\Var}{\mathrm{Var}\,}

\makeatletter
\newcommand*{\rightharpoonupfill@}{%
  \arrowfill@\relbar\relbar\rightharpoonup
}

\newcommand*{\leftharpoondownfill@}{%
  \arrowfill@\leftharpoondown\relbar\relbar
}

\newcommand{\xrightleftharpoons}[2][]{%
  \ensuremath{%
    \mathrel{%
      \settoheight{\dimen@}{\raise 2pt\hbox{$\rightharpoonup$}}%
      \setlength{\dimen@}{-\dimen@}%
      \edef\CA@temp{\the\dimen@}%
      \settoheight\dimen@{$\rightleftharpoons$}%
      \addtolength{\dimen@}{\CA@temp}%
      \raisebox{\dimen@}{%
        \rlap{%
          \raisebox{2pt}{%
            $%
            \ext@arrow 0359\rightharpoonupfill@{\hphantom{#1}}{#2}%
            $%
          }%
        }%
        \hbox{%
          $%
          \ext@arrow 3095\leftharpoondownfill@{#1}{\hphantom{#2}}%
          $%
        }%
      }%
    }%
  }%
}

\begin{document}

\title{Variance-corrected Michaelis-Menten equation predicts transient rates
of single-enzyme reactions and response times in bacterial gene-regulation}

\author{Otto Pulkkinen}
\affiliation{Department of Physics, Tampere University of Technology, FI-33101
Tampere, Finland}
\author{Ralf Metzler$^*$}
\affiliation{Institute for Physics \& Astronomy, University of Potsdam,
D-14476 Potsdam-Golm, Germany}
\affiliation{Department of Physics, Tampere University of Technology, FI-33101
Tampere, Finland}

\begin{abstract}
Many chemical reactions in biological cells occur at very low concentrations of
constituent molecules. Thus, transcriptional gene-regulation is often controlled
by poorly expressed transcription-factors, such as \emph{E.coli} lac repressor
with few tens of copies. Here we study the effects of inherent concentration
fluctuations of substrate-molecules on the seminal Michaelis-Menten scheme of
biochemical reactions. We present a universal correction to the Michaelis-Menten
equation for the reaction-rates. The relevance and validity of this correction for
enzymatic reactions and intracellular gene-regulation is demonstrated.
Our analytical theory and simulation results confirm that the proposed
variance-corrected Michaelis-Menten equation predicts the rate of
reactions with remarkable accuracy even in the presence of large non-equilibrium
concentration fluctuations. The major advantage of our approach is that it involves
only the mean and variance of the substrate-molecule concentration. Our theory is
therefore accessible to experiments and not specific to the exact source of the
concentration fluctuations.
\end{abstract}

\maketitle

The basic question of enzymology concerns the rate of a reaction, in
which a substrate-molecule $S$ first forms a complex $SE$ with an enzyme, and
upon catalysis turns into a product $P$. The reaction, commonly written as 
\begin{equation}
\label{mmscheme}
S+E\xrightleftharpoons[\mathit{k_{-1}}]{\mathit{k_1}} SE \xrightarrow{k_2}P+E,
\end{equation}
was first described and analyzed by Henri \cite{Henri1902,Henri1903}. His work
was picked up ten years later by Leonor Michaelis and Maud Leonora Menten, who
then presented a thorough derivation and interpretation of the equation
\cite{MM1913} 
\begin{equation}
\label{mme}
v (\rho)  = \frac{v_{max} \rho}{K_M +\rho}
\end{equation} 
for the reaction-rate $v$ as a function of the substrate concentration
$\rho$. $K_M=(k_{-1}+k_2)/k_1$ is the Michaelis-Menten constant, and the
maximal rate is $v_{max} = k_2\times[E]$, where $[E]$ is the enzyme concentration.
Equation \eqref{mme} is commonly known as the Michaelis-Menten equation (MME).

The derivation of the MME relies on a series of assumptions. First, the step in
which the SE complex is turned into a product, is in general reversible. This
can lead to 'blocking' of the reaction pathway by products turning back
into substrate-molecules, or even activate the reaction if the enzyme has
several conformational forms \cite{Popova1975, Lomholt2007}. However, if the
products are immediately removed from the system by some other reaction,
or their concentration is otherwise small, the backward reaction can be
neglected. The second major assumption in the derivation of the MME is that of
a quasi-steady state, which says that the concentration of the complex $SE$
in the reaction scheme \eqref{mmscheme} does not change considerably on the
time scale of the product formation \cite{BriggsHaldane}. Alternatively, one
may assume fast equilibration of the complex with the free substrate, which leads
to a slightly different form of $K_M=k_{-1}/k_1$ called the dissociation
constant in this case. The validity of the quasi-steady state approximation has
been discussed, for instance, by Rao and Arkin \cite{Rao2003}.

Third, and perhaps most fundamentally, the whole derivation of the MME is
based on deterministic ordinary differential equations, in which the total
substrate and enzyme concentrations enter as parameters. In practice, this
means that the solution of constituent molecules must be well-mixed by fast
diffusion to avoid local concentration differences. Moreover, there must be
large numbers of constituent particles in a large reaction volume in order
to be able to define concentrations and to neglect the fluctuations due to
the inherently
stochastic nature of the reaction and substrate import or synthesis. The
breakdown of the MME due to the finiteness of the reaction volume were predicted
and discussed by Grima \cite{Grima2009}. Also, the effects of small enzyme
numbers on the reaction have been under theoretical study since the 1960's
\cite{Bartholomay1962}, yet the statistics of the turnover times of single
enzymes were thoroughly analyzed \cite{Kou2005} and confirmed in the laboratory
only within the past decade \cite{English2006, Moffitt2010}. Theoretical results for
rates of single-enzyme reactions were reviewed in Ref.~\cite{Grima2014}.

In this article, we show that concentration fluctuations due to stochastic
substrate production and degradation can drastically change the
reaction-rate from the prediction of Eq.~\eqref{mme}, especially in the
particular problem of gene-regulation by highly specific transcription-factors.
Previously, St\'efanini, McKane and Newman \cite{Stefanini2005}
solved a model with input of single substrate-molecules into a small
cellular compartment which acts as the reaction volume. Their formula for
the reaction-rate is remarkable because it is a very simple function of
the mean compartmental substrate concentration alone, even if the process
effects a non-zero variance. This is probably due to the Poissonian nature
of the input process in their model. Already in the 1970's, Smeach and
Smith \cite{Smeach1972} had presented a solution to a more complicated
model with Poissonian removal of substrate-molecules. The solution is much
more complicated than the one by St\'efanini {\it et al.}, and cannot be
expressed solely in terms of mean substrate concentration. Effects of substrate
concentration fluctuations on turnover times of single-enzyme molecules were
discussed, for instance, in Refs.~\cite{Qian2002, Molski2008}. However,
no explicit, universal correction to Michaelis-Menten formula due to
substrate fluctuations appears to exist in the literature. In the next section,
we derive mathematically exact upper and lower bounds for the reaction-rates
of enzymatic reactions under such fluctuations and propose a first order
correction to the Michaelis-Menten equation. As we will demonstrate, this
variance correction works exceptionally well.

\section*{RESULTS}

\subsection*{Variance-corrected Michaelis-Menten equation}

It is a quite common feature that the concentrations of molecules that are
genetically expressed change in time in an abrupt manner: The concentration can
be almost constant for a long period of time, although the cell growth
causes some gradual change. Occasionally the concentration jumps by a
significant amount due to a translational (and/or transcriptional) burst
or cell division \cite{McAdams1997,Thattai2001, Swain2002, Paulsson2005, Kaern2005,
Toner2013,Lloyd-Price2014}. This behavior contrasts the slow
concentration drifts and only small fluctuations around the mean when there
is a large number of molecules present in a reaction volume such as a
biological the cell. For sufficiently large concentrations,
the rate of a reaction involving these molecules as a substrate follows the
MME {\eqref{mme}} with the constant concentration $\rho$ replaced by the
(time or ensemble averaged) mean $\langle \rho(t) \rangle$ of a stochastic,
time dependent concentration $\rho(t)$. On the other hand, in case of abrupt
and relatively large concentration changes, as in bursty protein production,
the reaction-rate $v(\rho(t))$, defined as the number of reactions
per unit of time, depends on the exact concentration $\rho(t)$ in each cell and
therefore shows high cell-to-cell variability. More precisely,
out of an experiment with a large sample of cells, the number
of reactions that have occurred up to time $t$ for each cell and the rate
computed from that information exhibits large cell-to-cell variations.
The quantity that describes the occurrence of the reaction on the
population level, and hence also provides the best estimate for the reaction
rate in a single cell, is the {\it ensemble average} $\langle v(\rho(t) )
\rangle$, {\it i.e.} the rate averaged over all cells. Next we show that this
estimate can be significantly lower than the naive approximation $v(\langle
\rho(t) \rangle)$ based alone on the mean substrate concentration in the
population.

The starting point for our analysis is the observation that the MME \eqref{mme}
is a concave function of the concentration $\rho$.
In terms of the normalized reaction-rate $v_0=v/v_{max}$, we
obtain by Jensen's inequality \cite{Kallenberg} for concave functions that 
\begin{equation}
\langle v_0(\rho(t))\rangle=\Big\langle\frac{\rho(t)}{K_M +\rho(t)}\Big\rangle
\leq\frac{\langle\rho(t)\rangle}{K_M+\langle\rho(t)\rangle}=v_0(\langle\rho(t)
\rangle).
\end{equation}
This means that the MME {\eqref{mme}} provides an upper bound for the reaction
rate for fluctuating substrate concentrations. Conversely, evaluating the
remainder in the integral form \cite{Apostol67} of Taylor's theorem, 
\begin{eqnarray}
v_0(\rho)&=&v_0(\rho_0)+v_0'(\rho_0)(\rho-\rho_0)\nonumber\\
&&+\int_0^1(1-x)v_0''(\rho_0+x(\rho-\rho_0))(\rho-\rho_0)^2dx,
\end{eqnarray}
with $\rho_0=\langle\rho(t)\rangle$, and after ensemble averaging we find
\begin{eqnarray}
\langle v_0(\rho(t))\rangle&=&v_0(\langle\rho(t)\rangle)-\frac{K_M}{(K_M+\langle
\rho(t)\rangle)^2}\Big\langle\frac{(\rho(t)-\langle\rho(t)\rangle)^2}{K_M+\rho(t)}
\Big\rangle\nonumber\\ 
\label{eq:mmlower}
&\geq& v_0(\langle\rho(t)\rangle)-\frac{1}{(K_M+\langle\rho(t)\rangle)^2}\Var
\rho(t),
\end{eqnarray}
where $\Var\rho(t)=\langle\left(\rho(t)-\langle\rho(t)\rangle\right)^2\rangle$
is the variance of the concentration $\rho(t)$, and the last
inequality follows from $\rho(t) \geq 0$. The inequality is clearly the
sharpest that can be expressed as a function of mean and variance of the
substrate concentration alone. Any further refinement requires the knowledge
of higher moments of the concentration distribution. As the MME \eqref{mme}
sets the optimal upper bound, given deterministic dynamics, we have
thereby accurately bounded the effects of stochastic substrate fluctuations
on the rate of the Michaelis-Menten reaction scheme \eqref{mmscheme} from
both above and below. The bounding of the Michaelis-Menten reaction-rate
is our first important result.

\begin{figure}
\centerline{\includegraphics[width=16cm]{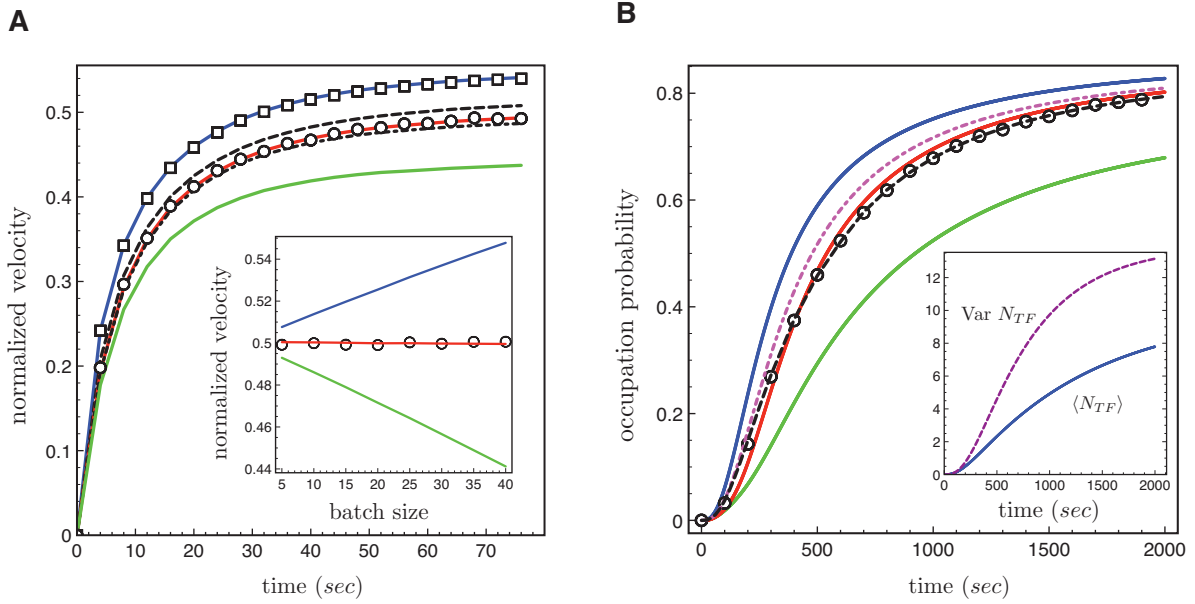}}
\caption{{\bf Stochastic corrections to rates of enzymatic reactions and
efficiency of transcriptional gene-regulation.} {\bf A} Normalized rate
$v_0$ of a single-enzyme reaction as a function of time for bursty substrate
production. The system is initially void of substrate-molecules. They are
produced in batches of size 40 with the production times determined by a
Poisson process, and there are on average 117.6 molecules present in the
stationary state. The variance-corrected MME (red solid line) coincides almost
exactly with the simulation results (circles). The uppermost, blue solid line is
the Michaelis-Menten upper bound based on the mean density alone, and the
lowermost, green solid line shows the optimal lower bound. The figure also
shows the third (dashed) and fourth (dash-dot) order Taylor approximations
to $\langle v_0 \rangle$. Squares are the solution of Stefanini {\it et al.}
\cite{Stefanini2005}
for the system with batch size 1. Inset: Predictions by the usual MME and
optimal lower bound deviate linearly from the variance-corrected MME and
simulation results with increasing batch size. {\bf B} The occupation probability 
$p_0$ of the TF-Operator complex as a function of time in a non-stationary gene
regulation experiment. The uppermost, blue solid line is the Michaelis-Menten
upper bound based on the mean density alone, and the lowermost, green solid
line shows the optimal lower bound. The variance-corrected MME (red solid
line) yields an excellent approximaxxtion to the results
\eqref{pbound}-\eqref{homogeneous} of the analytical model
(dashed line) and the simulation results (circles). The pink dash-dotted
line is the Poissonian approximation given by Eq.~\eqref{PoissonAppr}. Inset: 
The density (blue solid line) and population variance (purple dashed line) of
the number of TF molecules in a single cell as a function of time.}
\label{enzgenefig}
\end{figure}

The natural candidate for a refined equation for the reaction-rate is the one
given by a second order Taylor approximation for small fluctuations around the
mean. We call this the variance-corrected Michaelis-Menten equation (VCMME);
\begin{equation}
\label{varmm}
\langle v_0(\rho(t))\rangle\approx\frac{\langle\rho(t)\rangle}{K_M+\langle\rho(t)
\rangle}-\frac{K_M}{(K_M+\langle\rho(t)\rangle)^3}\Var\rho(t).
\end{equation}
It has the desired properties of reducing to the MME in the deterministic limit and
falling between the optimal variance bounds for all values of the mean concentration,
its variance and the constant $K_M$. In fact, the corrected equation \eqref{varmm}
attains
the mean value of the optimal upper and lower bounds at the mean concentration
$\langle \rho(t) \rangle= K_M$, which marks the point at which the usual MME
reaches half of its maximum and the enzymes are in the substrate-bound complex
state approximately half of the time. Fig.~\ref{enzgenefig} A shows that the
variance-corrected approximation is able to predict the rate of a reaction
catalyzed by a single enzyme in a cellular compartment with bursty input
of substrate-molecules: The only difference to the usual reaction scheme
{\eqref{mmscheme}} in this example is that the substrate-molecules enter
the reaction volume in batches. With growing batch size, the concentration
fluctuations increase and the contribution of the variance correction in
Eq.~{\eqref{varmm}} becomes more significant.

There are clear reasons why we truncate the Taylor series at the term proportional
to the variance: First, the truncated form is experimentally accessible. One only
needs to measure the substrate concentration variance in addition to the mean
concentration.
Second, the full Taylor series of the ensemble-averaged reaction rate
\begin{equation}
\langle v_0(\rho(t))\rangle = v_0(\langle  \rho(t) \rangle) -
\sum_{n=2}^{\infty}(-1)^{n}\frac{K_M\langle(\rho(t)-\langle\rho(t)\rangle)^n
\rangle}{(K_M+\langle\rho(t)\rangle)^{n+1}}, 
\end{equation}
is typically an alternating series (that is, successive terms have opposite
signs) because the right tail of the concentration distribution is longer and
also possibly heavier at low mean substrate concentrations in cellular
environments. It turns out that the third order approximation (series truncated
so that the last term is $n=3$) and fourth order approximation ($n=4$) can yield
poorer results than the series
truncated at the variance correction ($n=2$), as indicated by the simulation
results of Fig.~\ref{enzgenefig} A. In fact, nothing guarantees that the
rate of convergence of the Taylor series is fast, and a very high order
approximation might be needed to outperform the VCMME. In such cases, most
of the higher order terms eventually cancel. We conjecture that only if the
fluctuations are very large compared to the mean concentration and $K_M$, or
if the concentration distribution is highly skewed, higher order corrections
are needed to get an accurate estimate for the reaction-rate. This point is
further discussed below in an application to gene-regulation.

The VCMME works equally well for a system of multiple
enzymes, although the resulting higher rate of production and removal
of substrate from the system must be compensated by an increased rate of substrate
import, which reduces the relative strength of stochastic effects. Also,
the formula does not differentiate between different sources of noise that show up in
the variance. It is quite remarkable that the accuracy of the new formula
is insensitive to such details. It is even more surprising in the light of
the fact that the fluctuations of the substrate concentration depend on the
reaction-rate itself. This is because the reaction is, in absence of cell
growth, cell division and substrate efflux, the only pathway of substrate concentration dilution. 
Exact evaluation of the reaction-rate would therefore involve solving
a closed system of Kolmogorov equations as pursued, for instance, by Smeach and Smith
\cite{Smeach1972} and St\'efanini {\it et al.} \cite{Stefanini2005}
in the case of substrate input one molecule at a time. The assumption of
quasi-equilibrium permits bypassing the feedback loop and directly taking the
average of the reaction-rates over a distribution of substrate concentrations.
The VCMEE and the demonstration of its remarkably accuracy compared to fully
stochastic simulations is the central result of this work.

\begin{figure}
\centerline{\includegraphics[height=10cm,angle=90]{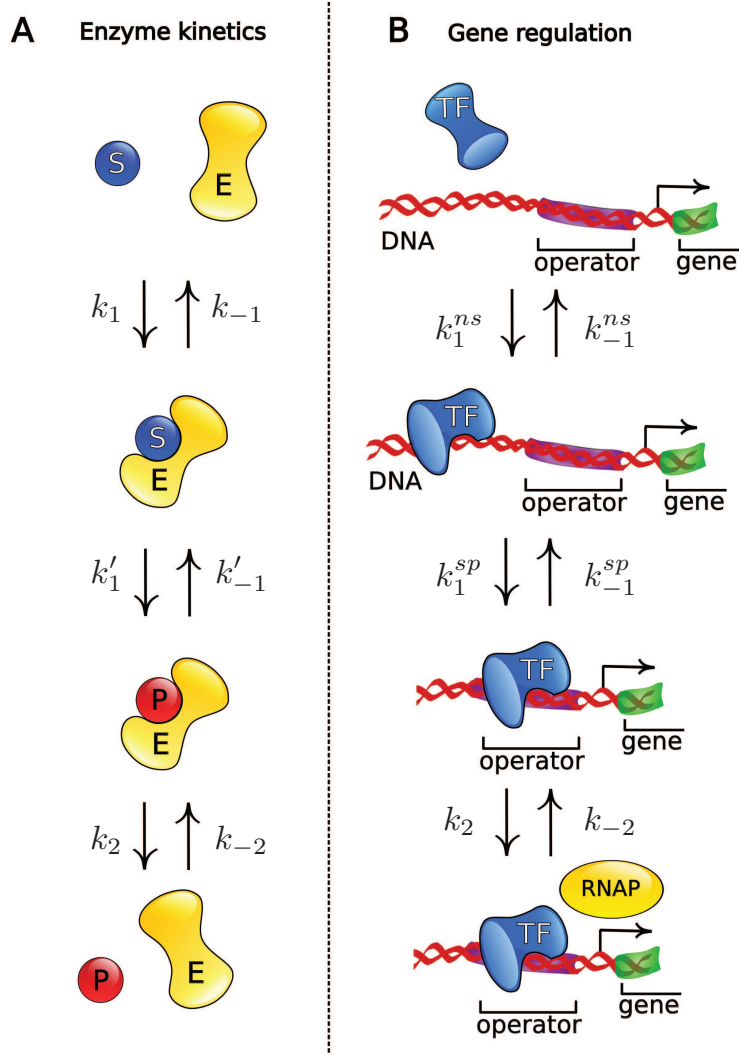}}
\caption{{\bf Similarities and differences in enzyme kinetics and
transcriptional gene-regulation.} {\bf A} In an enzymatic reaction, substrate
S and enzyme E form a complex SE, which turns into a complex PE of product
and enzyme. The product P is released in the final step. As a consequence,
the number of substrate-molecules in the reservoir is decreased by one. {\bf B}
In transcriptional gene-regulation, a TF molecule first binds the DNA
non-specifically and finds its specific binding site, the operator O, by
sliding along the DNA. The operator bound TF changes the DNA conformation or
interacts with RNA polymerase (RNAP) directly, which changes the transcription
rate of the target gene. The TF molecule is returned to the reservoir of
free TF upon unbinding from the DNA, and therefore the total number of TF
is conserved. The probability of the complex states (third step) both in
enzyme kinetics and gene-regulation is described by a Michaelis-Menten type
equation as discussed in the main text.}
\label{mmgenecartoon}
\end{figure}

\subsection*{Application to gene-regulation}

Transcription-factors (TFs) find their specific binding sites on the DNA by
facilitated diffusion: They diffuse in the cytoplasm and, upon contact, bind
non-specifically to the DNA and slide along it and simultaneously probe the
nucleotide sequence \cite{BWvH1982,BvH1989,slutsky, Gowers2005}. The enhancement of the
stochastic search of TFs for their specific binding site on the DNA was shown
to be relevant in living cells \cite{elf,spakowitz,plos,Bauer2015}. As shown in
Fig.~\ref{mmgenecartoon}, this process of non-specific binding of a TF to
the DNA and consequent specific binding to the operator is similar to the
formation of a substrate-enzyme complex and its transformation to a complex of
a product and enzyme. More precisely, the kinetics of a TF within a reaction
distance for non-specific binding to the DNA around the target corresponds
to an augmented Michaelis-Menten reaction without release of a product
(see the reaction scheme {\eqref{mmscheme}})
\begin{equation}
TF+DNA\xrightleftharpoons[\mathit{k_{-1}^{ns}}]{\mathit{k_1^{ns}}}TF\,DNA
\xrightleftharpoons[\mathit{k_{-1}^{sp}}]{\mathit{k_1^{sp}}}TF\,O.
\end{equation} 
Here $TF\,DNA$ represents the non-specifically DNA-bound TF and $ TF\,O$ is the
TF-operator complex.

As the TF slides along the DNA, the rate of interconversion between the
non-specific and specific binding modes of a TF must be very high ({\it e.g.}
order of $10^6/s$ for {\it LacI} \cite{BWvH1982}) for the TF to recognize
its binding site as it diffuses over it. Hence we assume equilibrium of
the non-specifically bound state. Denoting the equilibrium constants for
non-specific and specific binding by $K_{NS} = k_1^{ns}/k_{-1}^{ns}$ and
$K_{SP} = k_1^{sp}/k_{-1}^{sp}$, respectively, the occupation probability
$p_0$ of the operator as a function of the TF concentration $\rho(t)$
around the target can then be shown to obey a Michaelis-Menten type formula
\cite{pulme2013}
\begin{equation} 
\label{genemmeq}
p_0(\rho(t))=\frac{K_{NS}K_{SP}\rho(t)}{1+K_{NS}(1+K_{SP})\rho(t)}\approx\frac{
\rho(t)}{\tilde{K}_M+\rho(t)} 
\end{equation}
where $\tilde{K}_M=1/(K_{NS}K_{SP})$ and the last approximation is justified if
the binding at the operator is strong, that is, $K_{SP}\gg1$. This is typically
fulfilled for TFs \cite{pvh,hwa,stormo,febs}. We adopt this approximation in the
following analysis, yet the cases with smaller $K_{SP}$ can be treated in a
similar manner. We believe that the mapping of the TF binding process onto
the Michaelis-Menten scheme is an important concept in the study of molecular
signaling in biological cells.

In order to obtain a formula for the mean occupancy $\langle p_0 (\rho(t))
\rangle$ of the operator, we need the statistics of the TF concentration
$\rho(t)$, and for that a realistic stochastic model for the expression of
the TF. We include the necessary stochastic steps in TF production, given
by the coupled reactions
\begin{eqnarray}
\label{TFprod}
DNA+RNAP&\xrightarrow{a_{TF}(t)}& mRNA\xrightarrow{\gamma_m}\emptyset
\nonumber\\
mRNA&\xrightarrow{v_{TF}}&P^1\\
P^1\xrightarrow{\delta_1}P^2\xrightarrow{\delta_2}\ \ldots\!&\xrightarrow{
\delta_{\kappa-1}}&P^{\kappa}\xrightarrow{\delta_{\kappa}}TF\nonumber.
\end{eqnarray} 
The first line describes the synthesis of mRNA as the RNA polymerase binds to the
DNA, followed by the degradation of individual mRNA molecules by Ribonuclease E. The
second line represents translation of mRNA to polypeptide chains of amino
acids by ribosomes. The third line is the protein maturation process,
which involves $\kappa$ separate steps, such as formation of the nucleus and
secondary and tertiary structures in protein folding. Each step in the diagram
is modeled as a Poisson process with the indicated rates. The sum of the random
lag times $\Delta_i=1/\delta_i$ in protein maturation is denoted by $\Delta$.

The above process determines the number of mature TF molecules produced on
a given time interval. This data still needs to be converted to the concentration
$\rho$ of TF molecules available for binding at the target site, and the
effects of cell growth and division, as well as a possible efflux of molecules,
have to be included. We set
\begin{equation}
\label{TFconceq}
\rho(t)=\int_0^t\phi( t-s ) dN_{synth}(s),
\end{equation}
where $\phi(t)=\phi({\bf x}_{TF}, {\bf x}_{O}, t)$ is the concentration of TFs in
the reaction volume surrounding the operator (located at ${\bf x}_{O}$)
of the transcription unit given that a mature TF molecule emerged at
location ${\bf x}_{TF}$ at time zero. In particular, causality requires
$\phi(t) = 0$ for $t<0$. The stochastic process
$N_{synth}$ marks the TF synthesis times according to the reaction scheme
{\eqref{TFprod}}. In mathematical terms, $N_{synth}$ is a Poisson point
process with a stochastic varying intensity, the events of which are
each independently delayed by a random amount $\Delta$. More precisely, the
intensity of the Poisson process $N_{synth}$ is itself a random process $v_{TF}
N_{mRNA}(t)$, {\it i.e.}, the product of translation rate and the number of mRNA
molecules in the cell. The latter evolves according to a {\it birth-death
process} \cite{bdp} with mRNA synthesis rate $a_{TF}(t)$ (independent of the system
state) and degradation rate $\gamma_{m}$ for each mRNA molecule. As for
the enzymatic reactions discussed in the previous section, we only
consider the simplest time dependence of the transcription rate $a_{TF}(t)$
of the TF gene, namely a zero rate for $t<0$ and a constant rate $a_{TF}>0$
for $t\geq 0$. This implies the initial condition $N_{mRNA}(0) =0$ for the
number of TF transcripts. The reason for choosing this particular scenario
is that this way we can directly observe the response of the target gene
to a change in the transcription rate of its regulator.

The statistics of the TF available for binding can be solved exactly within
the theoretical framework explained above. This can be achieved by inspecting
the generating function for the master equation \cite{Shahrezaei2008} or,
alternatively, by using the Laplace transform $\langle \exp (-\lambda \rho)
\rangle$ of the TF concentration. The latter is convenient in our case because
the probability {\eqref{genemmeq}} of specific binding at the target can be
easily obtained by integration,
\begin{eqnarray}
\label{pbound}
\langle p_0(\rho(t))\rangle&=&1-\Big\langle\frac{\tilde{K}_M}{\tilde{K}_M+\rho(t)}
\Big\rangle\nonumber\\ 
&=&1-\tilde{K}_M\int_0^{\infty}e^{-\lambda\tilde{K}_M}\langle e^{-\lambda\rho(t)}
\rangle d\lambda. 
\end{eqnarray}
Given the definition {\eqref{TFconceq}} of the TF concentration we first observe
that the delay $\Delta=\Delta_1+\ldots+\Delta_{\kappa}$ for each point of the
process $N_{synth}$ has the distribution \cite{Feller2}
\begin{equation}
f_{\Delta}(t)=\sum_{i=1}^{\kappa}\delta_i e^{-\delta_i t}\prod_{\stackrel{j=1}{
j\neq i}}^{\kappa}\frac{\delta_j}{\delta_j-\delta_i} 
\end{equation}
for $t\geq 0$ and zero otherwise. Furthermore, the Laplace transform can be written
in the form (see SI for full derivation)
\begin{eqnarray}
\label{Lapleq1}
\langle e^{ -\lambda\rho(t)}\rangle&=&\langle\exp\big\lbrack-v_{TF}\left(N_{mRNA}
\ast F(\lambda,\cdot)\right)(t)\big\rbrack\rangle\\
\label{Fdef}
F(\lambda,t)&=&\left((1-e^{-\lambda\phi})\ast f_{\Delta}\right)(t),
\end{eqnarray}
with the asterisk denoting convolution $(f\ast g)(t)=\int_0^tf(s)g(t-s)\, ds$
starting at time zero. The angular brackets on the right-hand side of
Eq.~{\eqref{Lapleq1}} denote the ensemble average over trajectories of the
birth-death process $N_{mRNA}$ for mRNA synthesis. According to diagram
{\eqref{TFprod}}, the process has a state-independent birth rate $a_{TF}$
and death rate $\gamma_m$ per molecule. We show in SI that the
convolution functional of Eq.~{\eqref{Lapleq1}} can be calculated analytically
even for a general integrable function of time in place of $F (\lambda,
t )$ and therefore also for a general physical distribution $\phi({\bf x}_{TF}, {\bf
x}_{O}, t)$  of TF molecules in Eq.~\eqref{TFconceq}. Our final result for
the Laplace transform is
\begin{equation}
\label{Lapleq2}
\langle e^{-\lambda\rho(t)}\rangle=\exp\left(a_{TF}\int_0^t(1-\beta(\lambda,s,t))
ds\right)
\end{equation}
where
\begin{eqnarray}
\label{Lapleq3}
\lefteqn{\beta(\lambda,s,t)=\exp\left[-\int_s^t\left(\gamma_{\mathrm{m}}+v_{
\mathrm{TF}}F(\lambda,t-\tau)\right)d\tau\right]} 
\\
&&+\gamma_{\mathrm{m}}\int_s^t\exp\left[-\int_s^{\tau'}\left(\gamma_{\mathrm{m}}
+v_{\mathrm{TF}}F(\lambda,t-\tau)\right)d\tau\right]d\tau', \nonumber
\end{eqnarray}
and $F$ is as defined in Eq.~{\eqref{Fdef}}. The probability \eqref{pbound} of
specific TF binding at the target can now be computed by
substituting the exact form of the Laplace functional and integrating the
resulting expression numerically. Formulae \eqref{Fdef}, \eqref{Lapleq2} and \eqref{Lapleq3} 
together form another important result. 

At this point, we concentrate on the stochastic effects of small molecular copy numbers in
the cell, and hence leave the implications of the nucleoid and other structures
\cite{saiz,Thanbichler2005,wunderlich,Broek2008,Lomholt2009,olivier}, observed in
{\it E. coli} and other bacteria, as a subject of further study. Therefore
we take the TF transport to be a fast diffusive process in a homogeneous
environment, which can be modeled by setting
\begin{equation}
\label{homogeneous}
\phi(t) = \phi({\bf x}_{TF}, {\bf x}_{O}, t) = \frac{1}{V_{Cell}}e^{-\gamma_P t} \, ,
\end{equation}
where $\gamma_{P}$ is the rate of protein (in this case the TF) dilution
due to efflux, cell growth and division, and $V_{Cell}$ is the average
cell volume. In particular, the function $\phi$ does not depend on the
location of the TF gene or the binding site in the cell, whereas in a
more detailed model it would be a solution of a space-inhomogeneous diffusion
equation \cite{pulme2013,Cotrell2012}. Moreover, protein partition 
upon cell division is modeled by constant dilution of the TF concentration. 
Explicit inclusion of cell division events would increase the TF variability in the time trajectories from
individual cells \cite{Swain2002}, and although the consequent variability is
of different type than in gene expression noise, its effect can be small on a
population level \cite{Lloyd-Price2014}. Simulations on our system show that
the protein partitioning already averages out to a large degree in ensembles
that include data from hundreds of cells (see SI). 

Fig.~\ref{enzgenefig} B shows that for biologically realistic parameters the
analytical results \eqref{pbound}-\eqref{homogeneous} are in excellent
agreement with results from simulations using the Gillespie algorithm. The
VCMME provides a very good approximation to the exact result, whereas the
error in the prediction for the probability of specific binding given by the
ordinary MME can be as large as 0.15 and even larger for higher, plausible
TF translation rates (see SI for parameter sweeps). Moreover, the
difference is significant for the whole duration of the transient, extending
to half an hour in the example considered, but this time period can be even longer
in real cells with slowed-down TF transport and cell division effects. These results 
remain valid even if there are multiple TF binding sites on the DNA (see SI). We also
observe from Fig.~\ref{enzgenefig} B that the VCMME performs better than an
analytical model based on simple Poissonian fluctuations of the number of
TF molecules in the cell, in which case the occupation probability reads
\begin{equation}
\label{PoissonAppr}
\langle p_0(\rho(t))\rangle=1-\tilde{K}_M\int_0^1x^{\tilde{K}_M-1}e^{-\langle{
\rho(t)\rangle V_{Cell}}(1-x)}dx.
\end{equation} 
The last result means that it can be more important to include all sources
of noise in the first stochastic correction in the VCMME rather than try to come up
with an approximative model for which all the central moments can be computed.

Fig.~\ref{enzgenefig} B also shows that the curve for the binding probability
is a sigmoidal function of time---even in the case of the ordinary MME, which
itself is a
non-sigmoidal function of the TF concentration by definition. Consequently, the
initial slow rise must come from the time evolution of the mean concentration,
caused by the equilibration of the  transcription process and by
protein maturation delays. To this end, let us study the analytical formula
for the mean TF concentration,
\begin{equation}
\label{meanrho}
\langle\rho(t)\rangle=a_{TF}b_{TF}\left((1-e^{-\gamma_m\cdot})\ast f_{\Delta}\ast
\phi\right)(t),
\end{equation}
obtained by differentiating the Laplace transform {\eqref{Lapleq2}} with
respect to $\lambda$ and taking the limit $\lambda \to 0$. The new
constant $b_{TF}=v_{TF}/\gamma_m$ stands for the mean size of a translational
burst. Eq.~{\eqref{meanrho}} shows that the mean concentration is a convolution
of three terms, each of which represents a different process in the reaction
diagram {\eqref{TFprod}}: The exponential term describes the equilibration
of the number of transcripts, $f_{\Delta}$ protein maturation, and $\phi$
the distribution and dilution of mature TF. A concave function of time is
recovered only in the limit of infinitely fast translation and mRNA degradation
(in such a way that $b_{TF}$ approaches a non-zero constant) in the absence of
maturation delays. With delays, a simple calculation using definition
\eqref{meanrho} shows that the initial rate of growth is
\begin{equation}
\label{meanrhosmallt}
\langle \rho(t)\rangle\approx\frac{a_{TF}v_{TF}}{(\kappa+1)!V_{Cell}}\left(
\prod_{i=1}^{\kappa}\delta_i\right)t^{\kappa+1},
\end{equation}
so each stochastic intermediate step in protein synthesis increases the
exponent of time by one. Result {\eqref{meanrhosmallt}} can be used to
determine the number $\kappa$ of intermediate states in the maturation
process. The stationary mean concentration equals $a_{TF} b_{TF}/(\gamma_P
V)$ as expected \cite{Thattai2001}.

Similarly, the variance of the TF concentration can be obtained from the
second derivative of the Laplace transform {\eqref{Lapleq2}}. However,
it is not expressible as a simple convolution but takes the form
\begin{eqnarray}
\lefteqn{\Var\rho(t)=a_{TF}b_{TF}\big\lbrack\left((1-e^{-\gamma_m\cdot})\ast
f_{\Delta}\ast\phi^2\right)(t)}\\
&&\hspace*{-0.6cm}
+2b_{TF}\gamma_m\left([1-e^{-\gamma_m\cdot}]\ast\left([f_{\Delta}\ast\phi]
\times\left[e^{-\gamma_m\cdot}\ast f_{\Delta}\ast\phi\right]\right)\right)(t)
\big\rbrack. \nonumber
\end{eqnarray}
This is also a sigmoidal function of time---it only reduces to the known
concave limit as the maturation delays vanish and translational bursts are
instantaneous, that is, $\gamma_m,v_{TF}\to\infty$. The slight difference 
with the result of Refs.~\cite{Thattai2001,Paulsson2005}, as the homogenous 
diffusion kernel \eqref{homogeneous} is inserted, originates from the
dilution of proteins instead of explicit degradation. The graphs of the mean
concentration and variance as functions of time are depicted in the inset
of Fig.~\ref{enzgenefig}.

\begin{figure*}
\centerline{\includegraphics[width=16cm]{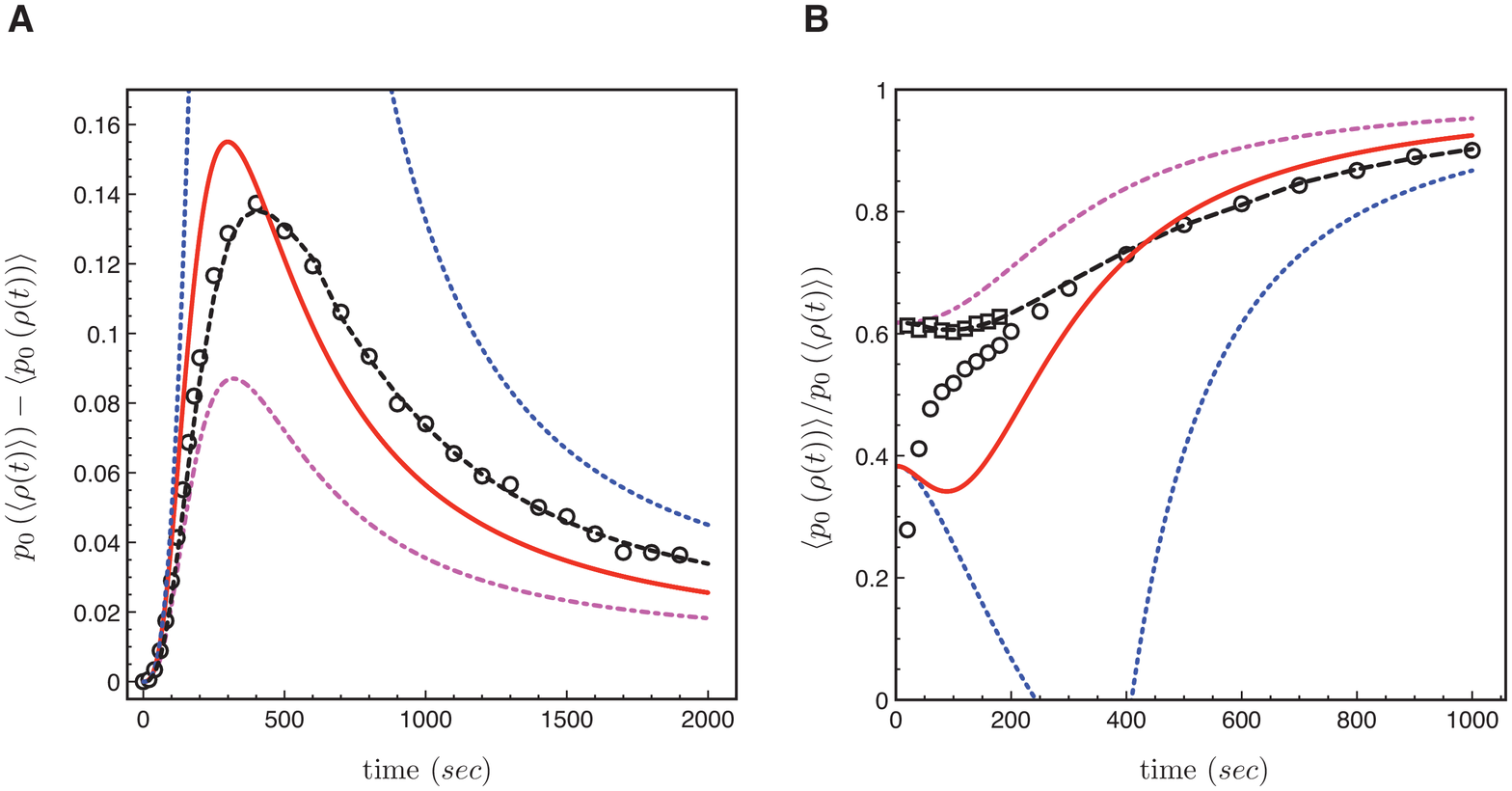}}
\caption{{\bf Sources and magnitude of stochastic corrections in
transcriptional regulation.} {\bf A} The curves show the difference of the
Michaelis-Menten prediction and various stochastic refinements for the
probability of specific binding at the operator. Circles are the Monte Carlo
simulation data, the black dashed line is the analytical model, the magenta
dash-dotted line is the Poisson approximation, and the red solid line is
the VCMME \eqref{varmm}.
{\bf B} Relative correction $\langle p_0 (\rho)
\rangle / p_0 (\langle \rho \rangle)$. The colour and line type coding is the
same as in the left panel. Additionally,  the squares are results of a Monte
Carlo simulation in which the TF is able to bind to the DNA in both nonspecific 
and specific forms independently. In that case, the TF-DNA complex is
close to equilibrium even at small times. Below 200 seconds, the Monte Carlo
simulation data of the full model differ significantly from the analytical
and numerical results of the model that assumes equilibrium of the TF-DNA
intermediate state.}
\label{generesultsfig}
\end{figure*}

Fig.~\ref{generesultsfig} {\bf A} shows that the error in the MME prediction for
the occupation probability is the largest at intermediate times. The curves
for the difference $p_0 (\langle \rho (t) \rangle ) - \langle p_0 (\rho(t))
\rangle$ share the same qualitative features in the full analytical,
variance-corrected and Poissonian models: The initial increase is well
approximated by $\Var \rho(t)/ \tilde{K}_M^2$, {\it i.e.} it is proportional
to the variance, as shown by the blue dashed line. The magnitude of the
correction reaches its maximum around 300 to 400 seconds, and finally, it
approaches a stationary value at rate $ \tilde{K}_M \Var \rho(t)/ \langle
\rho_t \rangle^3$.  The final value can be estimated from the stationary
mean and variance of the TF concentration using the VCMME. In the stationary
state, the variance of the TF concentration is proportional to the square
of the translational burst size. Since the stationary mean protein level is
not affected by the length of translation bursts, as long as the burst
size is constant, whereas the variance grows as the bursts get shorter,
the limit of instantaneous translation bursts provides an upper bound
for the magnitude of the variance correction to the MME. The maturation
delays also slightly increase the stationary variance---the effect being the
strongest for $\kappa=1$, {\it i.e.} for a single maturation step.  However,
since the stationary correction is small in comparison to the transient correction,
we associate the particularly strong deviation from the deterministic theory
with non-equilibrium fluctuations in the TF concentration, which arise at
short and intermediate times.

We stress that the TF concentration fluctuations are clearly the main reason for
the difference of the deterministic and stochastic occupation probabilities. In
particular, the assumption of fast equilibration of the TF-DNA complex
seems to have little or no effect at all. This is confirmed by the close
match of the analytical (assuming equilibrium) and simulation
(no equilibrium assumption made) results in Fig.~\ref{generesultsfig} {\bf
A}. The breakdown of an intermediate equilibrium can only be seen at the shortest
times by inspecting the ratio $\langle p_0(\rho(t)) \rangle /p_0(\langle
\rho(t) \rangle)$. In particular, the plot in Fig.~\ref{generesultsfig}
{\bf B} reveals that at the very beginning of the experiment and up to 200
seconds, when the TF concentration is still very small, the analytical model,
the VCMME, and the Monte Carlo simulation all exhibit different behavior:
The ratios from the analytical model and from the VCMME start at constant
levels at time zero, and have a minimum around 100 seconds, whereas the
stochastic simulation data increases monotonically. Extrapolation to
the beginning of the experiment suggests that the initial value in the
simulation is close to zero, implying a significant deviation from the
deterministic dynamics. Fig.~\ref{generesultsfig} {\bf B} also shows that
the results of the analytical model match the data from another Monte Carlo
simulation, in which the TF is able to bind to the DNA independently in
both the non-specific and specific forms, and binding kinetics therefore
equilibrates much faster. Hence, we associate the observed deviation to the
breakdown of the intermediate state equilibrium. The description of the TF
binding dynamics to DNA in terms of the VCMME relevant for typical in vivo
concentrations is our other central result.

\section*{DISCUSSION}

We demonstrated that stochastic concentration fluctuations can lead to
a significant correction to the Michaelis-Menten equation of reaction
kinetics. The proposed first order correction is conceptually simple and
experimentally accessible---it only requires the ensemble variance of the
fluctuating molecule concentration. Yet it turns out to be highly accurate
according to our analytical and numerical computations with detailed models
of catalyzed chemical reactions and transcriptional gene-regulation. In the
wake of massive experimental advances allowing single molecular insight into
signaling processes in living cells \cite{xie,hammart}
we believe that our results are both timely and relevant for a more accurate
description of reactions at typical in vivo concentrations.

Interestingly, a correction term to the MME resembling the one we propose, but
of a completely different origin, was derived in Ref.~\cite{Sinitsyn2010}. In
that analysis, a second term in the MME emerges
as a consequence of differentiable changes in \emph{deterministic} substrate
concentrations, and the only stochasticity in the problem lies in the kinetics
of the catalysis. A connection to non-equilibrium quantum mechanical geometric
phases was suggested. We believe that the similarity of the two theories can be
explained by the theory of linear response: The geometric phase correction
term to MME is proportional to the time derivative $\Dot\rho(t)$, which probably
can be expressed in terms of concentration fluctuations due to dilution
of substrate by the stochastic enzymatic reaction. However, the variance
correction to MME can be used to embrace other sources of stochasticity as
well---in particular upstream fluctuations, and it is not restricted to
small fluctuations and thus represents a more general and flexible concept.

It was recently suggested that the fluctuations of intermediate metabolite
concentrations in network structures consisting of multiple enzymatic processes
are typically uncorrelated \cite{Levine2007}. This would mean that spontaneous
fluctuations do not spread out in the network. In mathematical literature,
this is known to be a general feature of so called quasi-reversible stochastic
networks, for which a Poisson process input into the network implies a Poisson
process output \cite{Serfozo}. While this might be true in networks
consisting of enzymatic reactions only, we have seen that gene-regulation
networks are more susceptible to upstream fluctuations. In particular,
operator state fluctuations can determine much of the global state of
the network. Therefore, it is important to extend the study of fluctuation
propagation to larger scale networks using the corrections presented here. This
also requires a study of multimerization and cooperativity because the
Hill equation is not necessarily concave in those cases, and the rate at
mean concentration is no longer an upper bound for the true reaction-rate
\cite{Pirone2004}. However, the second order Taylor approximation is still the
natural candidate for a corrected equation of promoter activity in those cases.

Upstream fluctuations in substrate concentration and the kinetics of substrate
binding are not the only stochastic components in the reaction kinetics. For
example, conformational changes of enzymes \cite{Lomholt2007,Kou2005,Qian2002,
Min2006} may also affect the reaction-rate.  A recent study also pointed
out that the rate of an enzymatic reaction is sensitive to the form of
the distribution for the unbinding times of the substrate from the enzyme
\cite{Reuveni2014}. For example, in some cases, increasing the substrate
concentration only decreases the rate. Such unbinding distributions could be
a consequence of anomalous conformational fluctuations of substrate-molecules
and enzymes, as observed experimentally for single proteins \cite{Min2005a,
Min2005b}. The joint contribution of both conformational and concentration
fluctuations will potentially lead to unexpected phenomena in reaction
kinetics and needs to be thoroughly investigated.

We are confident that our results will inspire a series of new studies of the
detailed reaction mechanisms in enzymatic and gene-regulatory reactions at
typically small concentrations both in living cells and the extracellular
environment. A particularly pressing question is how such reactions are
modified in view of recent studies demonstrating the anomalous diffusion of
transcription factor-size green fluorescent proteins (GFPs) in the cytoplasm
and nucleoplasm of eukaryotic cells \cite{gratton} as well as in the presence of
superdiffusive, active mixing \cite{christine}.

\section*{METHODS}

All Monte Carlo simulations presented in this study were performed using the
Gillespie algorithm \cite{Gillespie}. The numerical integration of the analytical
results was implemented in Mathematica. The full list of parameters used in
producing the figures, as well as additional figures with higher transcription and
translation rates for the TF in gene-regulation,  can be found in the SI
Appendix. Derivation of formula \ref{Lapleq1} for the Laplace transform of
the TF concentration and a general formula for exponential functionals of
immigration-death processes, which leads to result \eqref{Lapleq2}, is also
provided in the SI Appendix.

\section*{Acknowledgements}

We acknowledge funding from the Academy of Finland (Suomen Akatemia) within the Finland Distinguished Professor (FiDiPro) program.

\section*{Author contributions}

O.P. and R.M. wrote the main manuscript text, O.P. prepared the figures, O.P.
and R.M. reviewed the manuscript.

\section*{Competing financial interests}

The authors declare no competing financial interests.

\end{document}